# Novel High Efficiency Quadruple Junction Solar Cell with Current Matching and Quantum Efficiency Simulations

Mohammad Jobayer Hossain[a], Bibek Tiwari[a], Indranil Bhattacharya[a]

[a] ECE Department, Tennessee Technological University, Cookeville, Tennessee, 38501, USA

*Abstract*: **A high theoretical efficiency of 47.2% was achieved by a novel combination of $In_{0.51}Ga_{0.49}P$, GaAs, $In_{0.24}Ga_{0.76}As$ and $In_{0.19}Ga_{0.81}Sb$ subcell layers in a quadruple junction solar cell simulation model. The electronic bandgap of these materials are 1.9eV, 1.42 eV, 1.08 eV and 0.55 eV respectively. This unique arrangement enables the cell absorb photons from ultraviolet to deep infrared wavelengths of the sunlight. Emitter and base thicknesses of the subcells and doping levels of the materials were optimized to maintain the same current in all the four junctions and to obtain the highest conversion efficiency. The short-circuit current density, open circuit voltage and fill factor of the solar cell are 14.7 mA/cm$^2$, 3.3731 V and 0.9553 respectively. In our design, we considered 1 sun, AM 1.5 global solar spectrum.**

## 1. Introduction

The inability of single junction solar cells in absorbing the whole solar spectrum efficiently and the losses occurred in their operation led the researchers to multijunction approach (Razykov et al., 2011; Xiong et al., 2010). A multijunction solar cell consists of several subcell layers (or junctions), each of which is channeled to absorb and convert a certain portion of the sunlight into electricity. Each subcell layer works as a filter, capturing photons of certain energy and channel the lower energy photons to the next layers in the tandem. The subcell layers are connected in series providing a higher voltage than single junction solar cells. Thus, utilizing the best photon to electricity conversion capability of each subcell, the overall efficiency of the cell is increased (Leite et al., 2013).

There are two methods of light distribution to the subcells in a multijunction cell. The first method uses a beam splitting filter to distribute sunlight to the series connected subcells and in the second method the subcells are mechanically stacked together (Imenes et al., 2004; Leite et al., 2013). The portion of the solar spectrum a subcell will absorb depends on the bandgap of the material used. Higher bandgap materials absorb higher energy photons and give relatively higher amount of voltage. Since number of higher energy photons is limited, number of excitons (electron-hole pair) generated and current is limited. On the contrary, materials with lower bandgap absorb lower to higher energy photons and give lower voltage but higher current. Therefore, choosing an appropriate set of high to low bandgap materials is important in multijunction solar cell design. This job can be challenging because the adjacent subcells should also be lattice matched to minimize threading dislocations (Patel, 2012).The presence of dislocation reduces the open circuit voltage ($V_{oc}$) and hence the overall conversion efficiency of the solar cell (Yamaguchi et al., 2005). Fortunately, there are some technologies that allow lattice mismatch up to certain limits. Metamorphic design uses buffers to limit formation of dislocations (King et al., 2007). Inverted metamorphic technology is a modified version of metamorphic technique where some cells are at first grown on a temporary parent substrate. The cells are then placed on the final substrate upside down and the temporary parent substrate is removed (Wanlass et al., 2015). Direct wafer bonding is another way which forms atomic bonds between two lattice mismatched materials at their interface and thus eliminates the dislocations (Moriceau et al., 2011). Some authors utilized this method successfully to address relatively higher mismatch value like 3.7% and 4.1% (Dimroth et al., 2014; Kopperschmidt et al. 1998).

After choosing the appropriate materials, current matching becomes the most important task in the design procedure. Since the subcells are connected in series, the lowest current density determines the overall current density of the cell. If current values are not matched, the excess current in the subcells other than the subcell with lowest current density gets lost as heat. The impact is twofold: firstly, some energy is lost; secondly, the heat generated deteriorates the cell performance further.

Solar cell is an excellent renewable power source. However, higher conversion efficiency and cost-effectiveness have been the major issues (Hossain et al., 2016). Theoretically a multijunction solar cell can provide 86.4% conversion efficiency with infinite number of junctions (Yamaguchi, Luque, 1999). Of course, manufacturing cost increases if higher numbers of junctions are used. When cost is an important factor in determining the market share of solar modules in the current power sector, we want to design a solar cell which has lesser number of junctions but gives relatively higher efficiency. The calculations using detailed balance method shows that, the highest efficiency achievable from a quadruple junction solar cell is 47.5% for single sun condition and 53% for maximum concentration of sunlight (Yamaguchi et al., 2004; King et al., 2009). This theoretical approach assumes ideal cases i.e. no reflection loss, zero series resistance of subcells and tunnel junctions, 300K temperature and no re-absorption of emitted photons (Leite et al., 2013). However, the highest practical efficiency achieved till now is only 46.0% which assembled four subcells with concentrators (http://www.nrel.gov). For 1-sun condition the efficiency is noticeably lower; 38.8% using five subcells (http://www.nrel.gov).

Solar energy ranges from ultraviolet to infrared region. Previously InGaP/GaAs/InGaAs (Wojtczuk et al., 2013) based triple junction solar cell was proposed which cannot capture much in the infrared region. To utilize infrared portions too, Ge was used as a bottom subcell layer (Yamaguchi et al. 2004; King et al., 2012; Green et al., 2015). Bhattacharya et al. proposed another material, InGaSb which is good at capturing infrared photons (Bhattacharya, Foo, 2013). It was used in GaP/InGaAs/InGaSb based triple junction solar cell later on (Bhattacharya, Foo, 2010; Tiwari et al., 2015; Tiwari et. al., 2016). In this paper, we propose an $In_{0.51}Ga_{0.49}P/GaAs/In_{0.24}Ga_{0.76}As/In_{0.19}Ga_{0.81}Sb$ based quadruple junction solar cell for the first time. The electronic bandgap of these materials are 1.9 eV, 1.42eV, 1.08 eV and 0.55 eV respectively which help proper distribution of light to all the junctions. The first two junctions are lattice matched. Lattice mismatch between GaAs and InGaAs is 2.78% when it is 5.59% between $In_{0.24}Ga_{0.76}As$ and $In_{0.19}Ga_{0.81}Sb$. Appropriate fabrication technique like metamorphic, inverted metamorphic or wafer bonding needs to be used to make the structure defect free. The simulation result shows that current density is same in all the junctions. This reduces the possibility of energy loss and performance deterioration. The theoretical efficiency of the cell is 47.2%. This value is higher than the present record efficiency quadruple junction solar cell with concentrators (46.0%) (http://www.nrel.gov).

## 2. Proposed Quadruple Junction Solar Cell

Material selection with proper bandgap is an important factor in designing high efficiency multijunction solar cell. III-V compound semiconductors are generally chosen because of their bandgap tunability through elemental composition. These compound semiconductor alloys have band gaps ranging from 0.3 to 2.3 eV which cover most of the solar spectrum (Leite et al., 2013). The proposed novel quadruple junction cell is also designed from III-V compounds, comprising InGaP, GaAs, InGaAs and InGaSb subcell layers respectively.

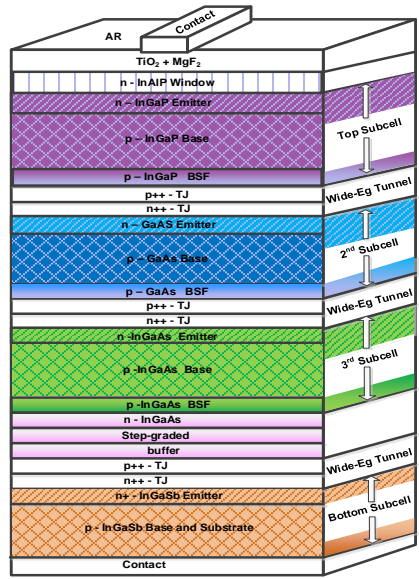

Fig.1.Structure of the novel quadruple junction solar cell

*A. Structure*

The quadruple junction solar cell consists of four subcells connected in series, as shown in Fig. 1. Each subcell has three parts: n type emitter, p type base and a back surface field (BSF) layer. Base is made thicker than emitter because of the work function of p type base being higher than n type emitter layer. The electron-hole pairs (excitons) are generated in the p-n junction formed in the interface between emitter and base which contributes to the photocurrent. The back surface field is made of the same material. It fixes dangling bonds and thus reduces surface recombination. Two adjacent subcells are connected together by tunnel diodes. Higher level of doping is used to design these tunnel diodes which help them not absorb light and exhibit tunneling effect. Antireflection (AR) coating is a special type of layer used to reduce reflection of light fallen on the solar cell (Saylan et al., 2015). With double layer $TiO_2$+$MgF_2$ antireflection coating, reflection loss can be reduced to 1%. The window layer acts as a means of light passage to the p-n junction. It protects the cell from outside hazards too. Step graded buffers are used to eliminate the threading dislocations formed between $In_{0.24}Ga_{0.76}As$ and $In_{0.19}Ga_{0.81}Sb$ lattice mismatched subcells. The front and back contacts are used to collect photocurrent from the solar cell.

*B. Material Properties*

The material properties considered for the design are summarized in Table I. Most of the properties are temperature dependent. All through the design process we considered 300K temperature. The top subcell is made of a high bandgap material, $In_{0.51}Ga_{0.49}P$ with bandgap of 1.9 eV (Schubert et al., 1995). This enables it to absorb photons in the ultraviolet region efficiently. GaAs has bandgap of 1.42 eV which empowers it to absorb most of the sunlight in visible range. The bandgap of $In_{1-x}Ga_xAs$ is $(0.36+0.63x+0.43x^2)$ eV (http://www.ioffe.rssi.ru). With x=0.76, it becomes 1.08 eV. The bottom subcell is made of low bandgap material, $In_{0.19}Ga_{0.81}Sb$ whose bandgap may be expressed as, $E_g$= $(0.7137-0.9445x+0.3974x^2)$ eV (Zierak et al., 1997), where x is the indium composition. With x=0.19, bandgap becomes 0.55 eV. Due to this lower bandgap value it can absorb in infrared region. The doping level of emitter is higher than base. Window layer is normally made of higher bandgap and highly doped n type material. Due to the high doping used and very little thickness, it does not absorb any photon and passes light to the subcells next in the tandem. The doping level of tunnel junction is even higher. The lattice constant of a material also depends on the composition. GaAs has a lattice constant of 5.65325 Å (http://www.ioffe.rssi.ru). The general expressions of lattice constants for $In_{1-x}Ga_xP$, $In_{1-x}Ga_xAs$ and $In_{1-x}Ga_xSb$ are $(5.8687-0.4182x)$ Å, $(6.0583-0.405x)$ Å and $(6.479-0.383x)$ Å respectively (http://www.ioffe.rssi.ru). The values become 5.653 Å, 5.8153 Å and 6.16 Å for $In_{0.51}Ga_{0.49}P$, $In_{0.24}Ga_{0.76}As$ and $In_{0.19}Ga_{0.81}Sb$ respectively. Since all these four materials have the same zinc blende crystal structure, defects occurred from the lattice mismatch can be easily eliminated by adopting appropriate technology i.e. metamorphic, inverted metamorphic, wafer bonding etc. Step graded buffers used in this structure solves the dislocation problem further. Minority carrier lifetime is another important parameter. If it is very low then some of the photocurrents are lost before they can be collected. It is in the order of $10^{-3}$ s for $In_{0.51}Ga_{0.49}P$ and $10^{-8}$ s for GaAs (Sun et al., 2015). For $In_{0.24}Ga_{0.76}As$, carrier lifetime depends on doping level through the relation, $\tau$= $(2.11*10^4 +1.43*10^{-10}*N+8.1*10^{-29}*N^2)^{-1}$ s (Ahrenkiel et al., 1998), where N is the doping density and $\tau$ is the carrier lifetime.

Front and back contacts are made of metals having very low resistances so that they can collect the generated photocurrent without any loss. The doping concentration for emitter of each subcell was designed to be in the order of $10^{18}$/$cm^3$. The highest value of doping concentration for base is in the order of $10^{17}$/$cm^3$. Surface recombination velocities of the materials used are in the order of $10^5$ cm/s (Thiagarajan et al., 1991; Boroditsky et al., 2000; Tanzid, Mohammedy, 2010). Therefore recombination losses were considered in the design.

TABLE I
MATERIAL PROPERTIES ASSUMED FOR THE DESIGN

| Material Properties | | Top subcell ($In_{0.51}Ga_{0.49}P$) | Subcell-2 (GaAs) | Subcell-3 ($In_{0.24}Ga_{0.76}As$) | Bottom Subcell ($In_{0.19}Ga_{0.81}Sb$) |
|---|---|---|---|---|---|
| Bandgap (eV) | | 1.9 | 1.42 | 1.08 | 0.55 |
| Lattice Constant (Å) | | 5.653 | 5.65325 | 5.8153 | 6.16 |
| Intrinsic Carrier Concentration (/$cm^3$) | | $1*10^3$ | $1.79*10^5$ | $1.31*10^9$ | $2.5*10^{13}$ |
| Surface Recombination velocity (cm/s) | | $4*10^5$ | $5*10^5$ | $1*10^4$ | $0.5*10^5$ |
| Dielectric Constant | | 11.8 | 12.9 | 13.3058 | 16 |
| Diffusion Coefficients | Electron | 26.8 | 200 | 220 | 297.7030 |
| | Hole | 3.8 | 0.5 | 0.09 | 0.5170 |
| Minority Carrier Lifetime (s) | Electron | $0.1*10^{-3}$ | $10^{-8}$ | $1.3562*10^{-7}$ | $9*10^{-9}$ |
| | Hole | $0.1*10^{-3}$ | $10^{-8}$ | $1.4149 \times 10^{-10}$ | $9*10^{-9}$ |
| Doping | Emitter | $8.5*10^{18}$ | $3.5*10^{18}$ | $8.5*10^{18}$ | $8.5*10^{18}$ |
| | Base | $3.5*10^{17}$ | $1.1*10^{15}$ | $5*10^{16}$ | $3.5*10^{17}$ |

To design the quadruple junction solar cell we made some assumptions that are generally done for simplification in solar cell modeling. These assumptions are (Kurtz et al., 1990): transparent tunnel junction interconnects with no resistance, no reflection loss and no series resistance loss in the junctions and p-n junctions formed are ideal (diode ideality factor, n is equal to 1).According to these assumptions, if a photon is absorbed by a subcell, one exciton (electron-hole pair) is generated. The fraction of the total number of photons absorbed in a subcell is determined by the thickness ($x_i$) of that subcell and the absorption coefficient ($\alpha$) of the constituent material.For our design, we collected the absorption data of In$_{.51}$Ga$_{0.49}$P, GaAs, In$_{0.24}$Ga$_{0.76}$As and In$_{0.19}$Ga$_{0.81}$Sb from Schubert et al.,1995, Adachi, 2009, http://www.ioffe.rssi.ru and Zierak et al., 1997 respectively.

We considered global AM 1.5 solar spectrum for photon flux incident on the solar cell. The top subcell absorbs a portion of this incident photon flux. The rest is transmitted to the next subcells. Thus, the photons incident on a subcell depends on the properties of the other subcells stacked above it in the tandem. If $\emptyset_s$ is the photon flux falling on the top subcell, the amount incident on any m$^{th}$ subcell lying below, $\emptyset_m(\lambda)$ can be expressed as equation (1). The percentage of absorbed photons converted into electron-hole pair in a subcell is called internal quantum efficiency (QE) of that subcell. It depends on absorption coefficient $\alpha(\lambda)$, base thickness $x_b$, emitter thickness $x_e$, depletion width $W$, base diffusion length $L_b$, emitter diffusion length $L_e$, surface recombination velocity in base $S_b$, surface recombination velocity in emitter $S_e$, base diffusion constant $D_b$ and emitter diffusion constant $D_e$, as given in equation (2) through (8).

$$\emptyset_m(\lambda) = \emptyset_s(\lambda) exp[-\sum_{i=1}^{m-1} \alpha_i(\lambda) x_i] \tag{1}$$

$$QE = QE_{emitter} + QE_{depl} + QE_{base} \times \exp(-\alpha(x_e + W)) \tag{2}$$

$$QE_{depl} = \exp(-\alpha x_e)[1 - \exp(-\alpha W)] \tag{3}$$

$$QE_{emitter} = f_\alpha(L_e) \left( \frac{l_e + \alpha L_e - \exp(-\alpha x_e) \times \left[ l_e \cosh\left(\frac{x_e}{L_e}\right) + \sinh\left(\frac{x_e}{L_e}\right) \right]}{l_e \sinh\left(\frac{x_e}{L_e}\right) + \cosh\left(\frac{x_e}{L_e}\right)} - \alpha L_e \exp(-\alpha x_e) \right) \tag{4}$$

$$QE_{base} = f_\alpha(L_b) \left( \alpha L_b - \frac{l_b \cosh\left(\frac{x_b}{L_b}\right) + \sinh\left(\frac{x_b}{L_b}\right) + (\alpha L_b - l_b) \exp(-\alpha x_b)}{l_b \sinh\left(\frac{x_b}{L_b}\right) + \cosh\left(\frac{x_b}{L_b}\right)} \right) \tag{5}$$

$$l_b = \frac{S_b L_b}{D_b} \tag{6} \qquad f_\alpha(L) = \frac{\alpha L}{(\alpha L)^2 - 1} \tag{7} \qquad l_e = \frac{S_e L_e}{D_e} \tag{8}$$

From equation (2), it is evident that the quantum efficiency of emitter, base and depletion region, all contribute to the overall quantum efficiency of the cell. Among the deciding factors of quantum efficiency, absorption coefficient, surface recombination velocity, diffusion length etc. are material properties which cannot be tuned once a particular material is chosen. However thickness can be easily varied in design process to obtain the highest possible quantum efficiency. The two quantities $x_e/L_e$ and $x_b/L_b$ are significant in the expression for emitter and base quantum efficiency. This gives an idea that the capability of tuning quantum efficiency by changing thickness is limited by the diffusion length of the material used. The value of $f_\alpha(L_e)$ and $f_\alpha(L_b)$ in equation 4 and 5 can be found from equation 7 by placing L= $L_e$ and L= $L_e$ respectively. Since we assumed no reflection loss due to the usage of double layer antireflection coating, internal quantum efficiency equals the external quantum efficiency. Changes in the quantum efficiency values with the change in thickness was investigated and illustrated in Fig. 2(a) through 2(d). We noticed that quantum efficiency increases with increase in base thickness up to a certain limit. After that, an increase in base thickness has no or little impact on quantum efficiency. Increase in emitter thickness on the contrary decreases quantum efficiency in most cases. The reason behind this is, work function for hole is greater than the electron. If emitter (n type) thickness increases, it absorbs some extra energy that would otherwise be absorbed in base (p type). Thus hole generation being impeded, quantum efficiency decreases.

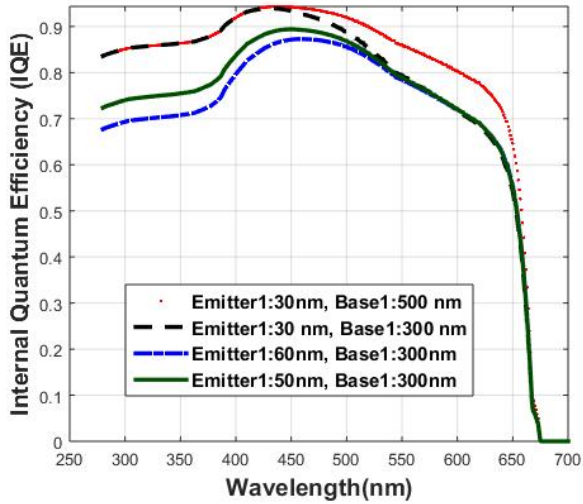

2(a) Top subcell (InGaP)

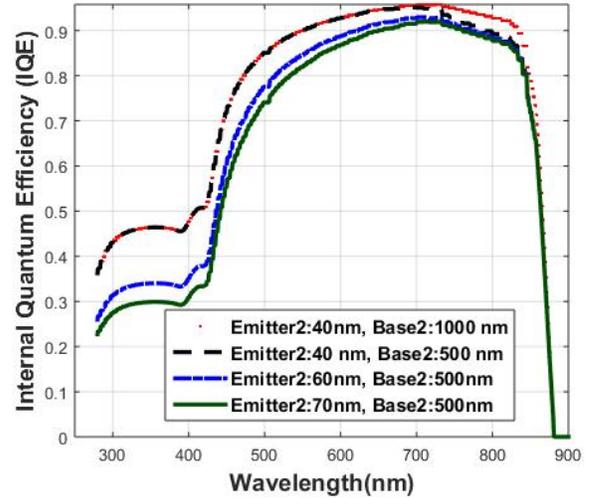

2(b) Second subcell (GaAs)

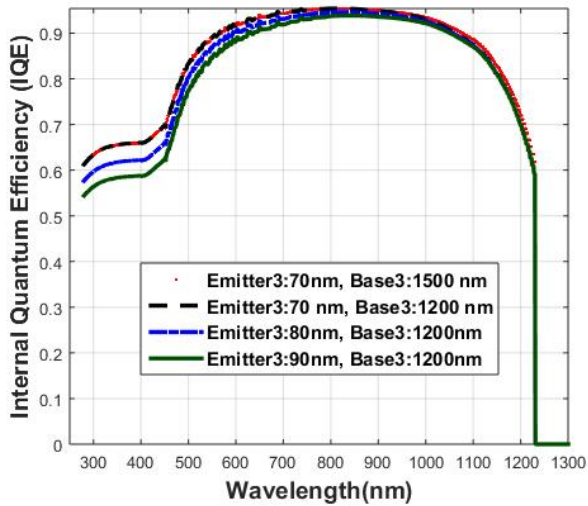

2(c) Third subcell (InGaAs)

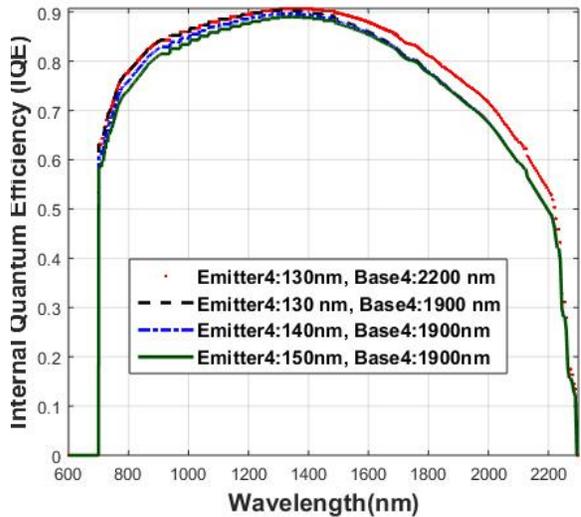

2(d) Bottom subcell (InGaSb)

Fig. 2. Change of quantum efficiency with change in thickness

The short circuit photocurrent density $J_{sc}$, obtained in a subcell depends on the quantum efficiency and the photon flux $\emptyset_{inc}$ incident on that subcell as follows,

$$J_{SC} = e \int_0^\infty \left( QE(\lambda) \Phi_{inc}(\lambda) \, d\lambda \right) \tag{9}$$

Here e is the charge of an electron ($1.6 \times 10^{-19}$ C). The incident photon flux $\emptyset_{inc}$ depends on the order of the subcell and geometry of the subcells above, as given in equation (1). In a solar cell, photocurrent is generated due to the minority electrons in the base and the minority holes in the emitter. Little amount of reverse current is also generated due to the majority carriers, which is a loss for solar cell. This current density is called dark current density ($J_0$). The photogenerated open circuit voltage can be written as,

$$V_{OC} \approx (kT/e)\ln(J_{SC}/J_o) \tag{10}$$

Where K is the Boltzmann constant and $T$ is the temperature in degree kelvin. Using the diode characteristic equation, we determine the effective photocurrent of a subcell as,

$$J = J_o\left[\exp\left(eV/nK_BT\right) - 1\right] - J_{SC} \tag{11}$$

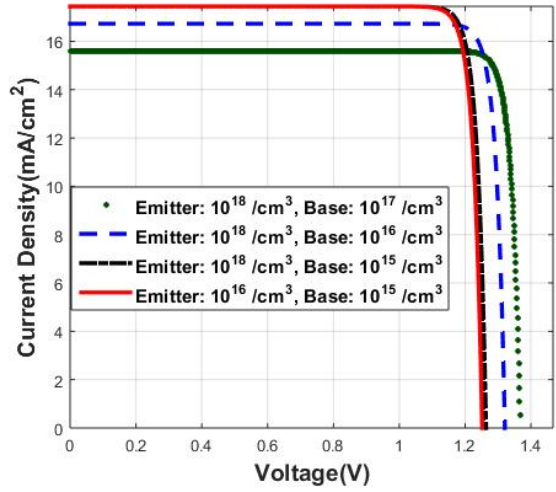

3(a) Top subcell (InGaP)

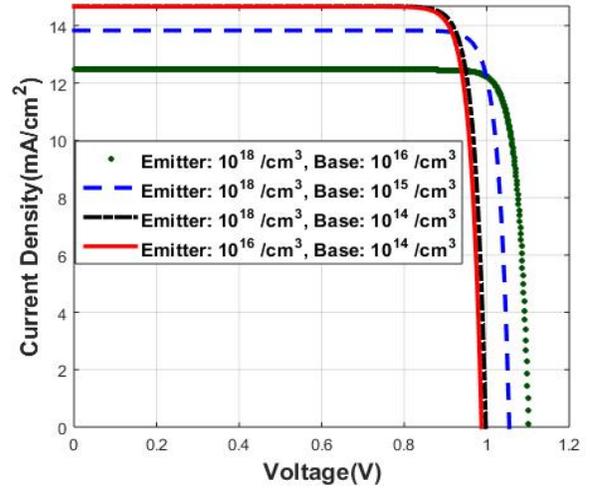

3(b) Second subcell (GaAs)

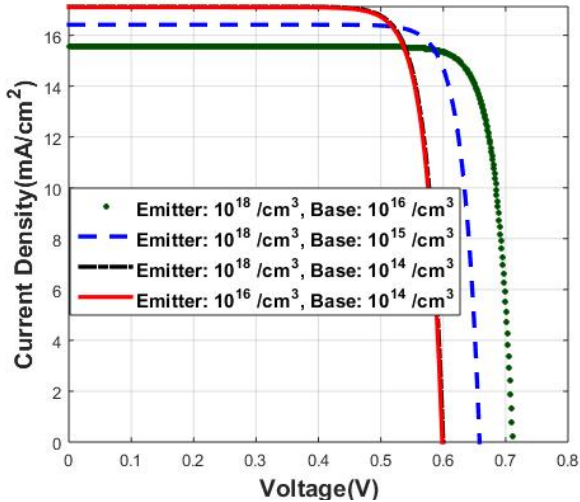

3(c) Third subcell (InGaAs)

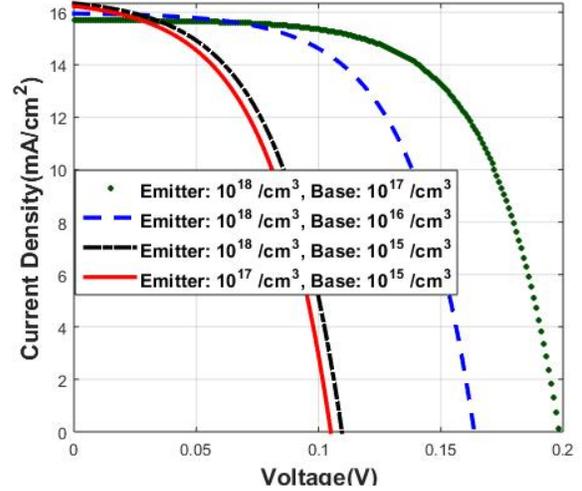

3(d) Bottom subcell (InGaSb)

Fig. 3. Change of J-V curve with change in doping

In our design approach, we have assumed each subcell as an ideal diode. Therefore diode ideality factor, n=1. Now, J-V characteristics of the subcells can be presented as change in current density (J) with respect to corresponding change in voltage obtained (V). These characteristics depend on material properties as well as design parameters like thicknesses and doping levels of the subcells. The impact of thickness variation on cell was discussed earlier in fig 2.

The impact of doping variation on current density and voltage produced is demonstrated in Fig. 3(a) through 3(d). All through this analysis, thicknesses of emitter for top, second, third and bottom subcells were kept unchanged at 30 nm, 40 nm, 70 nm and 130 nm respectively. The base thicknesses were 500 nm, 1000 nm, 1500 nm and 2200 nm respectively. As in Fig. 3(a), 3(b), 3(c) and 3(d), decrease in base doping decreases photovoltage and increases photocurrent. When emitter doping is decreased, it caused no or little decrease in voltage, the current being unchanged for the first three cases. For the bottom subcell, decrease in emitter doping decreases both voltage and current.

In a multijunction solar cell all the subcells are connected in series. Therefore, current matching is very important. If current density in all the subcells are not matched, the excess current in a subcell, being unable to flow, will be lost as heat. This thermalization will also give rise to deteriorated cell performance. In a current matched cell, the current density of the overall cell is the current density of any particular subcell, $J$. Also, the total open circuit voltage is the sum of the voltages in the subcells.

$$V_{total} = \sum_{i=1}^{m} V_i \qquad (12)$$

Fill factor for a solar cell can be empirically expressed as (Green, 1981),

$$FF = \frac{V_{OCnormalised} - \ln(V_{OCnormalised} + 0.72)}{V_{OCnormalised} + 1} \qquad \text{Where, } V_{OCnormalised} = \frac{e}{nkT} V_{OC} \qquad (13)$$

Finally, the conversion efficiency of a solar cell is,

$$\eta = \frac{J_{sc} \times V_{oc} \times FF}{P_{in}} \times 100\% \qquad (14)$$

Here $P_{in}$ is the input power (sunlight) to the solar cell. In standard test case it is 1000 W/m² for global AM 1.5 solar spectrum. The numerator expresses the power generated (electricity) from the solar cell per square meter. Doping and thickness value of the subcells were tuned to achieve the highest efficiency possible. The optimization trial is given in table 2.

TABLE II
DESIGN OPTIMIZATION FOR HIGHEST EFFICIENCY

| | Parameters | Design 1 | Design 2 | Design 3 | Optimized Design |
|---|---|---|---|---|---|
| Doping Density (/cm³) | Emitter 1 | 6.5×10¹⁷ | 8.5×10¹⁸ | 8.5×10¹⁸ | 8.5×10¹⁸ |
| | Base 1 | 3.5×10¹⁶ | 7.5×10¹⁶ | 7.5×10¹⁶ | 3.5×10¹⁷ |
| | Emitter 2 | 3.5×10¹⁷ | 3.5×10¹⁸ | 3.5×10¹⁸ | 3.5×10¹⁸ |
| | Base 2 | 0.1×10¹⁵ | 0.3×10¹⁵ | 0.4×10¹⁵ | 1.1×10¹⁵ |
| | Emitter 3 | 8.5×10¹⁷ | 8.5×10¹⁸ | 8.5×10¹⁸ | 8.5×10¹⁸ |
| | Base 3 | 0.2×10¹⁵ | 0.7×10¹⁵ | 0.7×10¹⁵ | 1.5×10¹⁶ |
| | Emitter 4 | 9.0×10¹⁷ | 9.0×10¹⁸ | 9.0×10¹⁸ | 8.5×10¹⁸ |
| | Base 4 | 8.5×10¹⁵ | 8.5×10¹⁶ | 8.5×10¹⁶ | 3.5×10¹⁷ |
| Thickness (nm) | Emitter 1 | 45 | 45 | 30 | 30 |
| | Base 1 | 220 | 300 | 270 | 400 |
| | Emitter 2 | 65 | 65 | 55 | 40 |
| | Base 2 | 700 | 900 | 700 | 1310 |
| | Emitter 3 | 95 | 90 | 70 | 70 |
| | Base 3 | 1540 | 1540 | 1460 | 1870 |
| | Emitter 4 | 150 | 150 | 120 | 140 |
| | Base 4 | 2820 | 2220 | 2200 | 2200 |
| Voltage (V) | Subcell 1 | 1.3313 | 1.3596 | 1.3579 | 1.4012 |
| | Subcell 2 | 0.9892 | 1.0236 | 1.0251 | 1.0663 |
| | Subcell 3 | 0.6152 | 0.6415 | 0.6413 | 0.6635 |
| | Subcell 4 | 0.1627 | 0.2173 | 0.2130 | 0.2422 |
| Matched Current, $J_{sc}$ (mA/cm²) | | 15.0 | 14.7 | 14.9 | 14.7 |
| Open Circuit Voltage, $V_{oc}$ (V) | | 3.0984 | 3.2420 | 3.2419 | 3.3731 |
| Fill Factor (FF) | | 0.9521 | 0.9538 | 0.9538 | 0.9553 |
| Efficiency (%) | | 44.1 | 45.3 | 46.0 | 47.2 |

In the first design all the base (p type) doping were kept in the order of $10^{17}/cm^3$ and emitter in the order of $10^{15}$ and $10^{16}$ /$cm^3$. Emitter thickness values were set to 45, 65, 95 and 150 nm for first, second, third and fourth subcell respectively. Base thicknesses were set to 220, 700, 1540 and 2820 nm respectively. With this arrangement, 44.1% conversion efficiency was found. Doping level was increased in the second design. Thickness values were also changed accordingly to match the short circuit current density at 14.7 mA/$cm^2$. This led to the increase of efficiency value to 45.3%. Thickness values were changed in design 3 keeping the doping level unchanged, except in base of subcell 2. With this trial open circuit voltage decreased little bit. However, the considerable increase in 2mA/$cm^2$ current contributed to the increased efficiency of 46.0 %. Finally, both doping and thickness values were tuned to different values. This step resulted in 47.2% efficiency with short circuit current density of 14.7 mA/$cm^2$, open circuit voltage of 3.3731 V and fill factor of 0.9553.

## 4. Analysis of the Optimized Design

### A. Quantum Efficiency

The cell was simulated to inspect its quantum efficiency and current density in each of the subcells. We considered global AM 1.5 solar spectrum for the simulation purpose. The internal quantum efficiency (IQE) plot in figure 4 clearly illustrates the absorption properties of the subcells as a function of wavelength. The top subcell, constructed from $In_{.51}Ga_{0.49}P$ showed good exciton (electron-hole pair) generation behavior in the higher frequency visible range. As in figure 4, its IQE was more than 90% for green light. GaAs subcell started absorbing when the top subcell was absorbing lesser number of photons. Its IQE was more than 90% in between 500 nm and 828 nm wavelength. It was placed below the top subcell in the stack so that the unabsorbed light can be absorbed by the second subcell. $In_{0.24}Ga_{0.76}As$ showed excellent IQE characteristics in a broad range. Note that, its IQE vale is comparable with GaAs in 500 nm - 828 nm range. If GaAs were not used in the second subcell, the generated current density through $In_{0.24}Ga_{0.76}As$ would be so high that current matching would be very difficult, resulting in lower cell efficiency. The bottom subcell, $In_{0.19}Ga_{0.81}Sb$ absorbed well in the infrared region unlike other subcells. The design ensured the right proportion of light distribution among all the subcells so that generated currents can be easily matched.

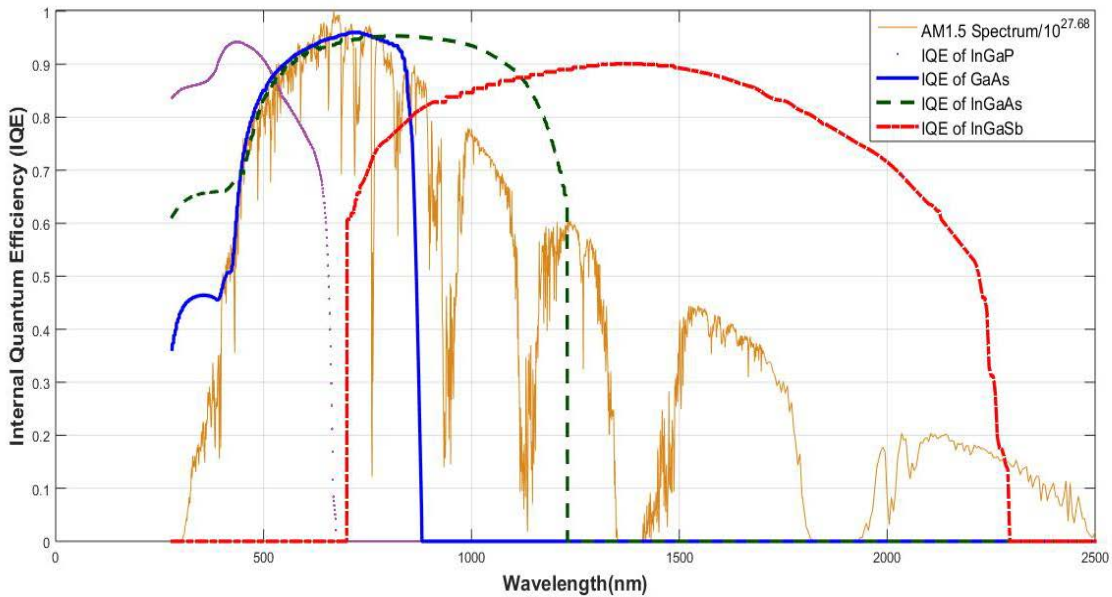

Fig. 4. Quantum efficiency plot of the individual junctions in the optimized design

### B. J-V Curve

Figure 5 gives an idea about the yield of the corresponding subcells. The top subcell generates the highest voltage

1.4 V with the lowest current of 14.7 mA per 1 cm² area. The bottom subcell on the contrary gives the lowest voltage (V) of 0.23 V with the highest current density (J) of 50 mA/cm². Second and third subcell followed this trend. Thicknesses of the subcells were tuned to attain current matching. In multijunction arrangement, the subcells are connected in series. Therefore, if current is not matched, the excessive current in a subcell, being unable to flow, would be lost as heat and the high temperature could harm the cell further. The J-V curve after current matching is shown in Fig 5(b).

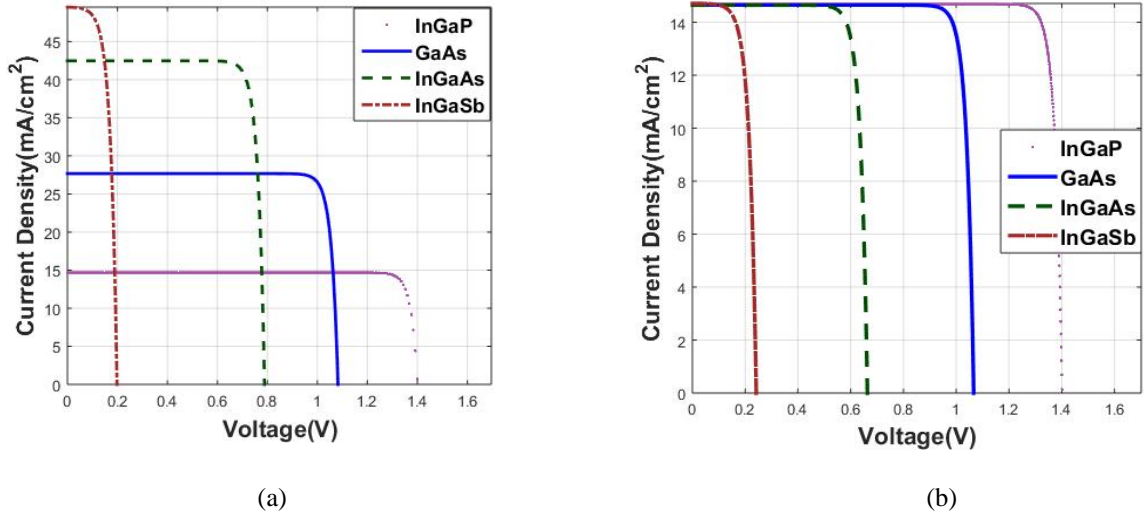

(a)          (b)

Fig. 5. J-V curve of the cell, (a) before current matching, (b) after current matching

### C. Non Ideal Diode

Previously we considered ideal diode with diode ideality factor if n=1. But in practical case this value is always greater than unity. Change in efficiency of the optimized design was inspected with variation in diode ideality factor value. As illustrated in Fig. 6, both the fill factor and efficiency decreases linearly with increase in ideality factor, n.

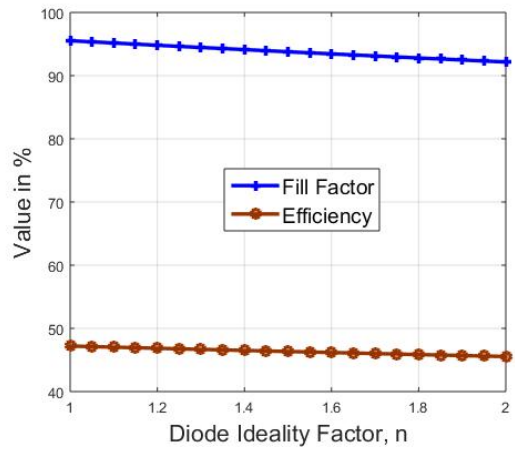

Fig. 6. Variation of fill factor and efficiency with diode ideality factor

In case of the best diode ideality factor, efficiency of the proposed optimized design is 47.2%. For a very bad junction diode with n=2, efficiency drops to 45.5%. This value is higher than the present record efficiency quadruple junction solar cell in single sun concentration (http://www.nrel.gov/).

## 5. Conclusion

A quadruple junction solar cell comprising $In_{0.51}Ga_{0.49}P$, GaAs, $In_{0.24}Ga_{0.76}As$ and $In_{0.19}Ga_{0.81}Sb$ subcell layers is proposed in this paper. This novel III-V combination gives high conversion efficiency of 47.2% for AM 1.5 global solar spectrum under 1 sun concentration. After careful consideration of important semiconductor parameters such as thicknesses of emitter and base layers, doping concentrations, minority carrier lifetimes and surface recombination velocities, an optimized quadruple junction design has been suggested. Current matching of the subcell layers was ensured to obtain maximum efficiency from the proposed design. Quantum efficiencies were subsequently determined for the matched current density of 14.7 mA/cm$^2$. The proposed quadruple junction solar cell is capable of absorbing and efficiently converting photons from ultraviolet to deep infrared region of the solar radiation spectrum.